\documentclass[%
 amsmath,amssymb,
 preprint,%
]{revtex4-1}

\usepackage{graphicx}
\usepackage{dcolumn}
\usepackage{bm}
\usepackage[utf8]{inputenc}
\usepackage[T1]{fontenc}
\usepackage{mathptmx}
\usepackage{etoolbox}
\usepackage{xcolor}
\newcommand\rev[1]{{\color{black}#1}}

\makeatletter
\def\@email#1#2{%
 \endgroup
 \patchcmd{\titleblock@produce}
  {\frontmatter@RRAPformat}{\frontmatter@RRAPformat{\produce@RRAP{*#1\href{mailto:#2}{#2}}}\frontmatter@RRAPformat}
  {}{}
}
\makeatother
\begin{document}

%\preprint{{\it Invited Contribution to JAP Special Issue: Two-Dimensional Materials and Heterostructures Under Strain}}

\title[]{Strength of 2D Glasses Explored by Machine-Learning Force Fields}
%Explored by Neural Network Force Fields

\author{Pengjie Shi}
\author{Zhiping Xu}
\email{xuzp@tsinghua.edu.cn}
\affiliation{ 
Applied Mechanics Laboratory, Department of Engineering Mechanics, Tsinghua University, Beijing 100084, China
}

\date{\today}

\begin{abstract}
The strengths of glasses are intricately linked to their atomic-level heterogeneity.
Atomistic simulations are frequently used to investigate the statistical physics of this relationship, compensating for the limited spatiotemporal resolution in experimental studies.
However, theoretical insights are limited by the complexity of glass structures and the accuracy of the interatomic potentials used in simulations.
Here, we investigate the strengths and fracture mechanisms of 2D silica, with all structural units accessible to direct experimental observation.
We develop a neural network force field for fracture (NN-F$^{3}$) based on the deep potential-smooth edition (DeepPot-SE) framework.
Representative atomic structures across crystals, nanocrystalline, paracrystalline, and continuous random network glasses are studied.
We find that the virials or bond lengths control the initialization of bond-breaking events, creating nanoscale voids in the vitreous network.
However, the voids do not necessarily lead to crack propagation due to a disorder-trapping effect, which is stronger than the lattice-trapping effect in a crystalline lattice and occurs over larger length and time scales.
Fracture initiation proceeds with void growth and coalescence, and advances through a bridging mechanism.
The fracture patterns are shaped by subsequent trapping and cleavage steps, often guided by voids forming ahead of the crack tip.
These heterogeneous processes result in atomically smooth facets in crystalline regions and rough, amorphous edges in the glassy phase.
These insights into 2D crystals and glasses, both sharing SiO$_{2}$ chemistry, highlight the pivotal role of atomic-level structures in determining fracture kinetics and path selection in materials.
\end{abstract}

\maketitle
\section{Introduction}
The strengths of crystals and glasses have garnered significant attention in statistical physics for their non-equilibrium nature\cite{nicolas2018deformation,bonamy2011failure,wondraczek2022advancing}, and are highly sensitive to the local arrangement of atoms\cite{font2022predicting,du2021predicting}.
In contrast to edge cleavage in crystals, failure of glasses often involves nucleation of voids in regions not necessarily at the crack tip.
The cracks can be trapped by the amorphous network, adding kinetic contributions to the energy cost of cracking, and the voids ahead of the crack tip could guide its advancement\cite{chang2024deep,hao2019atomistic}.
Recent studies show that nanostructured glasses in the form of nanocrystals\cite{fokin2006homogeneous,xu2024effect,liao2020real} and paracrystals\cite{tang2023toughening} can markedly enhance their fracture resistance.
Introducing nanocrystalline domains\cite{gallo2022fracture} increases fracture toughness by $80\%$, from $0.6$ to $1.1$ MPa m$^{1/2}$, while paracrystalline glasses exhibit a threefold greater enhancement\cite{tang2023toughening}.
It is thus interesting to explore the failure process of nanostructured glasses by considering their atomic-level structures, beyond the continuum framework\cite{shi2024non}.

Direct characterization of glass structures remains challenging\cite{zhigong_2021,feng2022experimentally,ly_2017,huang_sciadv_2020,huang_prl_2020}.
Structural description of glasses is limited to the short-range order (SRO)\cite{liao2020real} and mediate-range order (MRO) parameters\cite{tang2023toughening}.
Notably, two-dimensional (2D) silica has recently been discussed in the literature since its successful synthesis.
The crystalline and amorphous structures can be directly visualized by electron microscope (EM) \cite{buchner2014ultrathin,huang2013imaging,huang2012direct}.
Exploring 2D silica thus allows direct characterization of their failure behaviors at the atomic level from both theoretical exploration and experimental observations, which remain largely unexplored.

Atomistic simulations are powerful tools to reveal atomic-scale material kinetics\cite{buehler_2010,atrash2011evaluation}. 
The accuracy and efficiency of their prediction are limited by the fidelity of interatomic interaction models, from first-principles calculations at the electronic-structure level to empirical force fields.
Empirical models such as Sundarararaman-Huang-Ispas-Kob (SHIK)\cite{sundararaman2018new,sundararaman2019new}, Du\cite{du2015challenges}, Bertani-Menziani-Pedone (BMP)\cite{bertani2021improved} were developed for oxide glasses including silica and successfully applied to problems such as material discovery and mechanistic analysis\cite{pedone2022interatomic,deng2019molecular}.
However, the capability of modeling fracture is under question due to the presence of strong lattice distortion and dangling bonds at the crack tip and cleaved edges\cite{buehler2007threshold,feng2022experimentally,liu2010graphene,qu_2022}, factors typically not account for force field parametrization.

Recent advances in developing machine-learning force fields (MLFFs)\cite{friederich_2021} allow us to revisit the non-equilibrium processes of material failure\cite{shi2024non} by using chemically-accurate force fields trained in the Deep Potential-Smooth Edition (DeepPot-SE)\cite{NEURIPS2018_e2ad76f2}, Neural Equivariant Interatomic Potential (NequIP)\cite{anderson2022e}, Neuroevolution Potential (NEP) frameworks\cite{fan2022improving}.
The MLFF approach has recently been applied to fracture problems by expanding the training dataset to include intermediate structures during fracture, which are usually not well modeled by empirical force fields\cite{shi2023planning,shi2024non}.
In this work, we develop a neural network force field for fracture (NN-F$^{3}$) tailored for 2D silica across various nanostructures, including crystals and nanocrystalline (NCG), paracrystalline (Para), continuous random network (CRN) glasses.
We investigate their mechanical responses at the atomic level to gain insights into the kinetics of material failure.

\rev{\section{Methods}
\subsection{Density functional theory (DFT) calculations}}
To obtain the quantitative relations between energies, forces on atoms, and the atomic-level structures, spin-polarized DFT calculations are performed using the Spanish Initiative for Electronic Simulations with Thousands of Atoms (SIESTA) package\,\cite{siesta} with numerical atomic orbitals (NAOs) at the double-$\zeta$-plus-polarization (DZP) level\,\cite{fajardo2022carrier}.
Perdew-Burke-Ernzerhof (PBE) parameterization of the generalized gradient approximation (GGA) is used for the exchange-correlation functional\,\cite{PBE}.
Troulliere-Martins-type norm-conserving pseudopotentials are chosen for the ion-electron interactions\,\cite{troullier1991}. 
The cut-off energy for electron wave functions is $500$ Ry.
The $\mathbf{k}$-space is sampled by a Gamma-centered, $4\times3\times1$ Monkhorst-Pack grid for the $144$-atom model.
For structures with open edges, sampling at the same $\mathbf{k}$-point density is used.
These settings assure a convergence threshold of $1$ meV/atom.

\rev{\subsection{NN-F$^{3}$ Development}}
To develop the NN-F$^{3}$ model for 2D silica, we first extend our approach from crystalline to glassy silica.
Building upon our previous work on 2D crystals\cite{shi2023planning}, we first pre-sample the 3D space of basal-plane strain states using a low-accuracy empirical force field (Tersoff\cite{munetoh2007interatomic}).
The strain sweeping is conducted at a rate of $1\times10^{-5}$ ps$^{-1}$ under controlled conditions of $10$ K using a Nos\'e-Hoover thermostat with a damping constant of $1$ ps.
Structures generated from this pre-sampling process undergo subsequent DFT calculations.
We employ the end-to-end, symmetry-preserving DeepPot-SE framework\cite{NEURIPS2018_e2ad76f2} and DeePMD-kit\cite{deepmd} for training the model.
Despite these efforts, the NN-F$^3$ model trained on this initial dataset shows some inaccuracies.
Notably, stress predictions near peak strain and Poisson ratios at high strain deviate from the reference DFT calculations\cite{shi2023planning}, highlighting the constraints inherent in sampling atomic-level structures generated by the Tersoff force field.

To enhance predictions of atomic forces in highly distorted structures at crack tips and cleaved, undercoordinated edges, we expand the initial dataset.
Implementing an active learning strategy (\emph{Training-Exploration-Labeling}\cite{dpgen,shi2024non}), we iteratively explore the most relevant structures based on predefined criteria related to deviations in atomic forces.
Initially, four NN-F$^3$ models are trained using different seeds for random number generation during parameter initialization,\cite{dpgen}.
MD simulations using these models are conducted in the Atomic Simulation Environment (ASE)\cite{ase}, generating trajectories and computing atomic forces under identical loading conditions as the pre-sampling process.
The \emph{Query by Committee}\cite{smith2018less} algorithm is employed to screen atomic-level structures, selecting those with maximum standard deviations (SD) of atomic forces exceeding $0.1$ eV/\AA~among the four NN-F$^3$ models.
Structures containing crack tips and open edges were identified by atoms with coordination numbers below $2$ for O atoms.
Additionally, structures with an SD of atomic forces greater than $0.2$ eV/\AA~are selected.
These screened structures undergo DFT calculations for labeling, enriching the dataset.
Consequently, atomic-level structures containing crack tips are incorporated into the dataset following the MD exploration process.

For 2D glassy silica, we employ an active-learning strategy based on the NN-F$^3$ model previously trained for 2D crystalline silica.
Initially, we generate all $144$-atom inequivalent structures, each with non-isomorphic bonding networks, through bond rotation\cite{klemm2020silica}.
This process results in $257$ structures distributed across $34$ distinct ring distributions.
From these, we select $45$ structures for active learning across the full strain space.
Structures exhibiting a maximum SD of atomic forces exceeding $0.2$ eV/\AA~(from the four NN-F$^3$ models) are chosen for subsequent DFT calculations.
We conduct $30$ iterations of active learning to achieve convergence, resulting in a reduction of the maximum SD of atomic forces to below $0.2$ eV/\AA.
The final dataset comprises $264,420$ data frames, represented in a sketch map illustrating the diversity of atomic-level structures in the training dataset, encompassing lattices under varied strain states, cleaved edges, and structures containing crack tips (Fig. S1a).
While these $144$-atom structures are too small to distinguish NCG, Para, and CRN structures, they effectively capture the essential structural characteristics of both crystalline and glassy silica.

In training the DeepPot-SE framework, we set the sizes of the embedding and fitting networks to $(25, 50, 100)$ and $(240, 240, 240)$, respectively, with a cutoff radius of $8.0$~\AA~and a smoothing parameter of $1.0$~\AA.
The batch size is $8$, and the hyperparameters \emph{pref\_e} and \emph{pref\_f}, which determine the weights of energy and force losses in the total loss function, are set to $1.0$ and $10.0$, respectively.
We employ Adaptive Moment Estimation (Adam) optimization over $3\times 10^{6}$ batch steps, starting with a learning rate of $0.001$ that exponentially decays to $1.0 \times 10^{-8}$ by the end of training.
Ninety-five percent of our dataset is used for training, with the remaining five percent reserved for validation.

\rev{\subsection{MD simulations}}
To simulate the fracture of 2D silica, we use the large-scale atomic/molecular massively parallel simulator (LAMMPS)\,\cite{lammps}. 
The structures (NCG, Para, CRN) are simulated with a cubic box of $200^{3}$ \AA, which contains $\sim15,000$ atoms, $\sim4$ times larger than previous studies\,\cite{font2022predicting}.
The fracture tests were conducted using an athermal quasistatic (AQS) protocol. 
In AQS, the structure is deformed at a strain rate of $5 \times 10^{-5}$ per step, followed by damped dynamics with a viscous rate of $1$ ps$^{-1}$ to relax the structure under a force-on-atom threshold of ${10}^{-4}$ eV/\AA.
%To determine the nucleation of voids, Si-O bonds are considered broken if their lengths exceed $2.2$~\AA, which is significantly larger than half of the critical Si-O-Si distance ($\sim1.8$ \AA).

\section{Results and Discussion}
\subsection{Performance of NN-F$^{3}$ for 2D Silica}
The performance of NN-F$^{3}$ is validated for 2D silica crystals and glasses through the predicted energies, forces, and stress-stain relations.
The root mean square error (RMSE) of the energy per atom, the interatomic forces, and the in-plane stress are below $0.79$ meV/atom, $37.6$ meV/\AA, and $24$ N/m, respectively, for the validation dataset (Fig. S2a-c).
The RMSE of the phonon spectrum measured from the DFT results is below $0.2$ meV (Fig. S2d).
%for all crystals and glasses
The uniaxial stress-strain relations of 2D silica crystals and glasses with structures not in the 45 structures of 2D silica exhibit excellent agreement with the reference DFT calculation results (Figs. S2 and S3).
These results validate the predictions of equilibrium properties. 
To verify the accuracy of NN-F$^3$ in describing the non-equilibrium fracture process, we study fracture of crystalline silica using both NN-F$^3$ and DFT-based Born-Oppenheimer MD simulations.
Runs of $0.3$ ps and $0.4$ ps are conducted for the zigzag and armchair crack edges, respectively, revealing almost identical structural evolution in the NN-F$^{3}$ and DFT results (Fig.~S4).

\subsection{Glassy structures}

A critical issue in the study of non-crystalline materials is whether the structural models accurately reflect those observed in experiments, particularly concerning their non-equilibrium properties.
The complexities in the chemistry and atomic-level structures of glasses present a challenge in verifying whether structures generated by methods like bond swapping\cite{welch2022high} or simulated annealing accurately capture the essential structural features.
Theoretical models were proposed with a CRN of Si-O tetrahedra in silica and by asserting that alkali or alkaline earth metals are glass modifiers that disrupt the network\cite{varshneya2013fundamentals}.
However, a direct comparison with experimental evidence at the atomic level is difficult.
Consequently, the agreement is often assessed by the physical properties of the generated structures such as the density, modulus, and radial distribution functions (RDFs)\cite{liao2020real}.

2D amorphous materials such as glassy silica\cite{huang2013imaging,huang2012direct} and monolayer amorphous carbon (MAC)\cite{toh2020synthesis,chen2021stone} were observed under an EM, which allow for direct determination of the atomic positions.
Experimental evidence shows that compared to MAC featuring a sub-crystalline 2D structure and significant out-of-plane displacements\cite{el2022exploring}, amorphous silica exhibits a CRN structure and lacks out-of-plane fluctuation for its symmetric $3$-atom-layer structure\cite{font2022predicting}.

\begin{figure*}[h]% [\sidecaptionrelwidth]%[ht]%
{
\includegraphics[width=\textwidth]{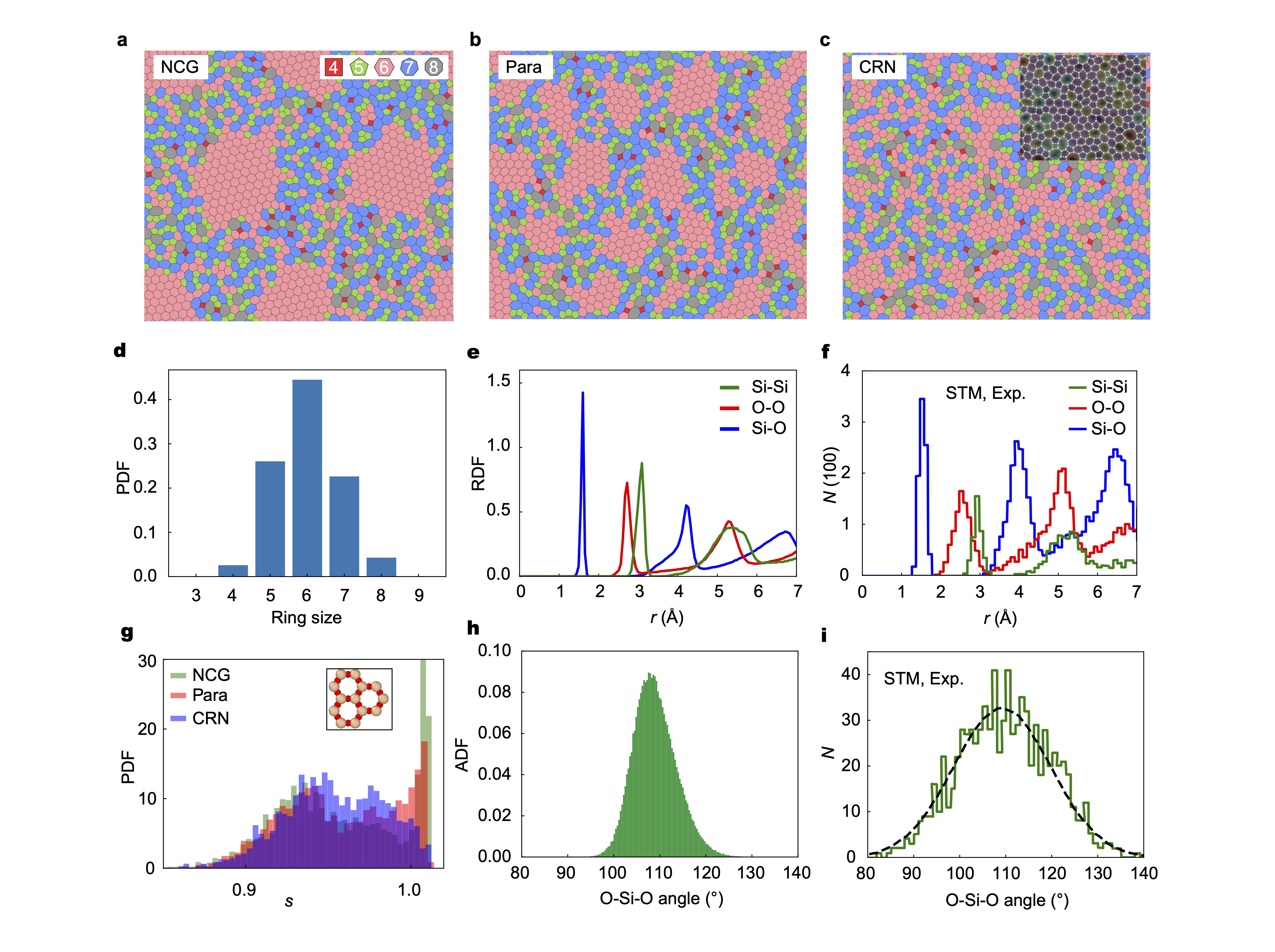}
\caption{}
\label{Fig_1}
}
\end{figure*}

2D glasses in a CRN structure can be generated by consecutive Stone-Wales (SW) rotations, where the topological disorder is controlled and measured through the ring statistics\cite{klemm2020silica}.
A Monte Carlo dual-switch procedure is deployed to the network of Si atoms\cite{morley2018controlling,font2022predicting}, the position of which is adjusted by minimizing a spring-like potential for ring-to-ring interaction in the dual space after each Monte Carlo move.
We use a fictitious temperature of $10^{-4}$ and $10^4$ Monte Carlo steps and a value of $\alpha=1/3$ for the Aboav-Weaire law\cite{vincze2004aboav}, which aligns with experimental evidence and suggests that small rings tend to sit around larger ones\cite{kumar2014ring}.
Structures of NCG and Para are constructed by preserving crystallites with sizes of $2.8$ nm and $1.4$ nm during structural transformation, respectively.
Computer-generated NCG, Para and CRN structures (Fig.\,\ref{Fig_1}a-c) exhibit identical ring distributions (Fig.\,\ref{Fig_1}d).
\rev{The glassy structures are rich in pentagons and heptagons, with rare quadrilaterals, which agrees well with experimental characterization using scanning tunneling microscopy (STM)\cite{buchner2014ultrathin}.
Their RDFs (Fig.\,\ref{Fig_1}e-f) and angle distribution functions (ADFs, Fig.\,\ref{Fig_1}h, i) are also similar and align well with the experimental data.
Minor discrepancies in ADF compared to experimental data can be attributed to two factors: the inaccuracy in determining atomic positions and the presence of structural defects within the materials.
Pinpointing exact atomic locations, particularly for oxygen atoms, remains challenging, often resulting in uncertainties observed in STM images\cite{buchner2014ultrathin}. Furthermore, defects are commonly overlooked in silica model construction, yet they significantly influence residual stress distribution and O-Si-O bond angles.
The RDFs and ADFs of Para, CRN, and NCG structures show minor differences (Fig. S5).
The RDF comparisons between CRN and Para structures align with previous findings\cite{tang2023toughening}.
Para structures notably feature a narrower angle distribution (SD $= 5.73^\circ$), compared to CRN (SD $= 6.32^\circ$) and NCG (SD $= 6.40^\circ$) structures, despite similar mean angles ($108.90^\circ$ for Para, $108.91^\circ$ for NCG, and $108.99^\circ$ for CRN, respectively).
These results suggest a higher degree of structural order in the Para structures, possibly indicative of medium-range order (MRO).}

To quantify the effect of nanostructuring on the MRO, a similarity order parameter $s$ between a cluster (A) and its reference (B) is defined as\cite{tang2021synthesis}

\begin{equation}
    s^{B} = \max_{l,\alpha}\frac{N_B\sum_i^{N_A}\sum_j^{N_B}\exp\left[-|\mathbf{T}\cdot r^A_i-r_j^B|^2/\sigma^2\right]}{N_A\sum_i^{N_B}\sum_j^{N_B}\exp\left[-|r_i^B-r_j^B|^2/\sigma^2\right]}.
\end{equation}

\noindent where $N_A$ and $N_B$ are the numbers of atoms in clusters A and B, respectively.
Here $\sigma=0.2$ \AA~is a smearing parameter. 
The transformation matrix $\mathbf{T}$ incorporates both affine scaling transformation ($l$) and cluster rotation ($\alpha$).
A crystalline cluster is chosen here as the reference (Fig.\,\ref{Fig_1}g, inset).
A distinct difference in the distributions of $s$ is identified, even though their RDFs, ADFs, and ring statistics are similar (Fig.\,\ref{Fig_1}g), suggesting the difference in lattice distortion.

\subsection{Void nucleation by bond breakage}
%Void nucleation in materials signifies the initiation of inelastic deformation, a critical aspect for engineering applications.
%Furthermore, exploring the potential connection between nucleation and fracture underscores the significance of this question.
Crack nucleation in crystals can be driven by phonon stability, which defines their strengths\cite{liu2007ab}.
In contrast, for glassy materials, the fate of embryo cracks is significantly influenced by the heterogeneity in both the chemistry and structures.
Understanding the link between material failure via bond breakage and structural features in the equilibrium state is a fundamental inquiry in the theory of glasses\cite{font2022predicting,dong2023non,cubuk2015identifying,fan2021predicting}.
We conduct AQS simulations under uniaxial tension along various directions by enforcing strain in the tensor form of
\begin{equation}
    \mathbf{E} = \left[\begin{matrix}
    \cos^2\theta&\frac{1}{2}\sin2\theta\\
    \frac{1}{2}\sin2\theta&\sin^2\theta,
\end{matrix}\right]\varepsilon
\end{equation}
where $\theta$ represents the tensile direction, and \rev{$\varepsilon$ is the amplitude of strain applied to the sample.}
Void nucleation occurs when the bond strength is exceeded\cite{zhang2022strength}, typically detected by the Si-O bond length surpassing $2.2$~\AA.
The nucleation strain is defined as the value at which the first bond breaks.
The size effect is examined by studying representative models with $144$ and $3,456$ atoms, corresponding to lateral sizes of $\sim 2$ nm and $\sim 10$ nm, respectively\,\cite{font2022predicting}.

\begin{figure*}[h]% [\sidecaptionrelwidth]%[ht]%
{
\includegraphics[width=\textwidth]{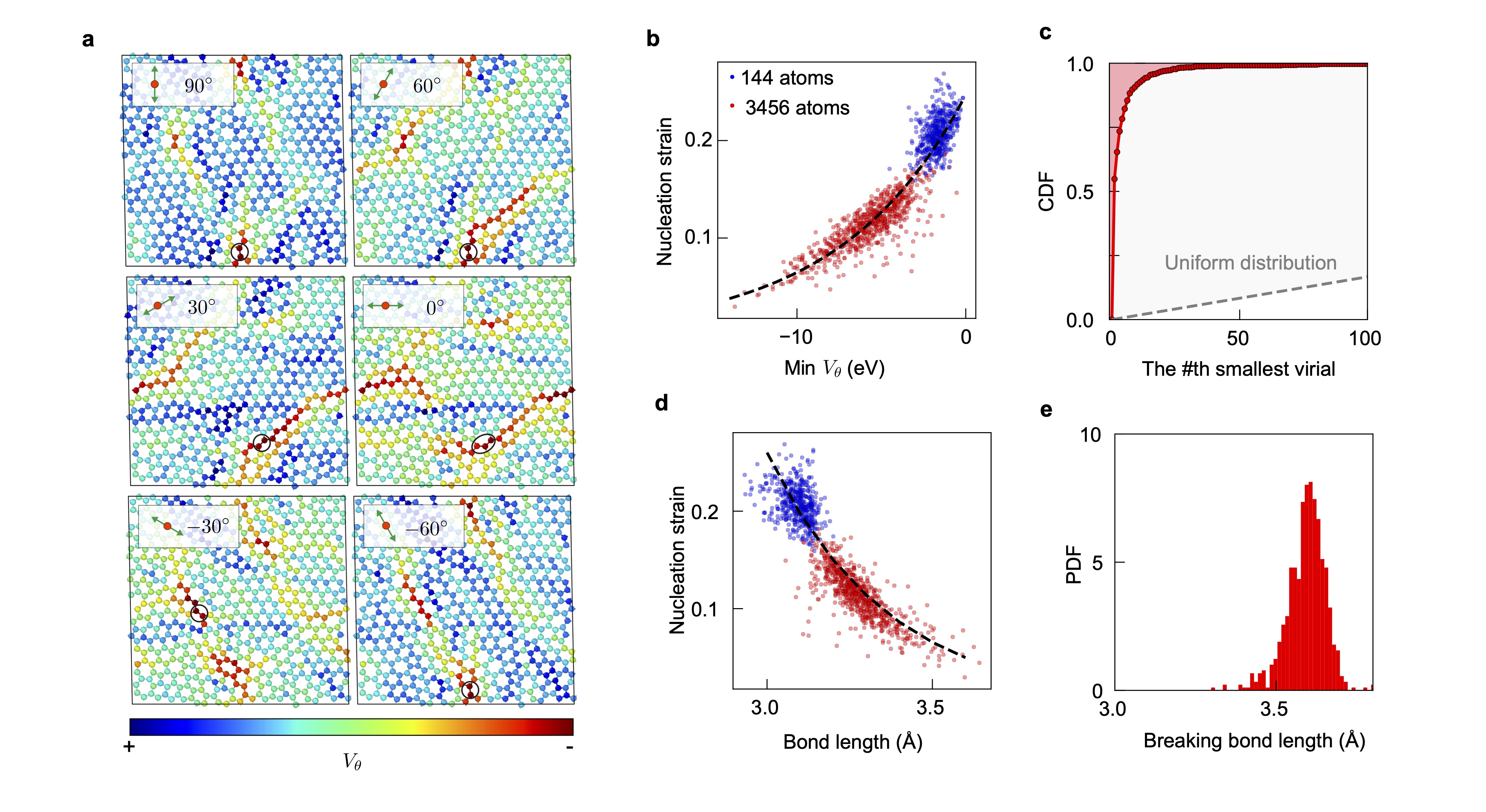}
\caption{}
\label{Fig_2}
}
\end{figure*}

\rev{Virial coefficients ($V_\theta = \mathbf{d}_{\theta}^\mathrm{T} \mathbf{V} \cdot \mathbf{d}_{\theta}$) for equilibrium, undeformed samples are computed along tensile directions ($\mathbf{d}{\theta}$), following the definition of normal stresses (Fig.~\ref{Fig_2}a).
\rev{The virial tensor, which relates to the stress tensor as $\mathbf{\sigma} = - \mathbf{V}/a$ ($a$ is the average area of atoms), characterizes residual stress induced by material heterogeneity.}
The spatial distribution of atomic sites exhibiting large $\lvert V_{\theta} \rvert$ values reveals the presence of `force chains' within the amorphous bonding network of the equilibrium or undeformed state.

There is a strong correlation between nucleation events and the value of $V_\theta$ (Fig.\,\ref{Fig_2}b), enabling the prediction of 2D glass strength from their equilibrium structures.
Voids typically nucleate at atomic sites with high residual stress.
Fig.\,\ref{Fig_2}c shows the cumulative probability density function (CDF) of $V_\theta$ at the sites of void nucleation, ranging from the smallest (the most stretched) to the largest (the most compressed) values.
Unlike a uniform distribution, the shape of CDF indicates that a few highly stretched states of atomic stress in equilibrium determine the strength of 2D glasses under tension.

On the other hand, the Si-O-Si bond lengths (distances between the silicon atoms) measuring the local strain also exhibit a strong correlation with the strain to void nucleation (Fig.~\ref{Fig_2}d).
The lengths of critical bonds that break in the damped dynamics follow a narrow distribution with a mean of $3.6$ \AA~and an SD of $0.07$ \AA, respectively~(Fig.~\ref{Fig_2}e).
%The reduction in the value of \pj{$U_{\rm n}$} or the virial/bond length descriptors with the size of models indicate the statistical nature.
The reduction in the values of virial/bond length descriptors with the size of models indicates the statistical nature (Figs.~\ref{Fig_2}b-d).
\rev{Heptagonal and octagonal rings are more prone to breakage than hexagonal rings, leading to more frequent void nucleation around larger rings (Fig. S6).}
These arguments made to CRN models apply to NCG and Para as well.}
 
\subsection{Fracture by void coalescence and bridging}
\begin{figure*}[h]% [\sidecaptionrelwidth]%[ht]%
{
\includegraphics[width=\textwidth]{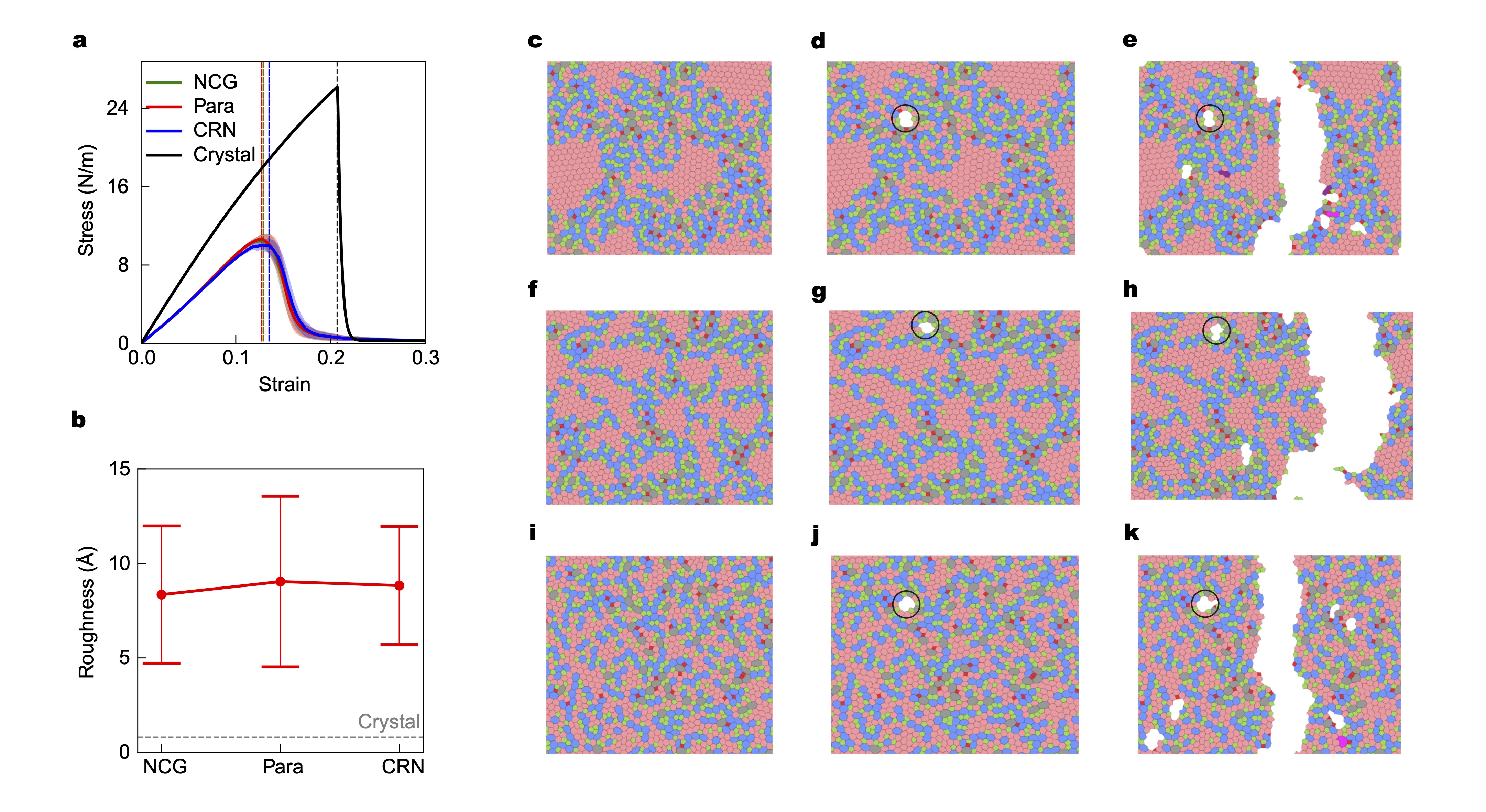}
\caption{}
\label{Fig_3}
}
\end{figure*}

The elastic responses of materials are lost after void nucleation.
Catastrophic fracture may be triggered by further increasing the loading amplitude.
% The loading rate is confirmed to be low enough to exclude the dynamical effects by comparing the simulation results to those at lower rates.
The strength of glasses is much reduced compared to the crystals, from $26.2$ N/m to $10$ N/m for CRN.
The stiffness decreases from $154.5$ N/m to $79.9$ N/m.
For the glassy phases, although the SROs and MROs (e.g., RDFs, ADFs, ring statistics) are close, the strength of Para and NCG structures increases slightly by $5\%$ and $4\%$ from CRN, respectively (Fig.\,\ref{Fig_3}a).
Significantly, the locations of fracture nucleation differ from those of void nucleation (Fig.\,\ref{Fig_3}c-k).
The CDF of the minimum distances between the sites of void nucleation and the crack suggests a weak correlation (Fig. S7).
This finding contrasts with crystals, where nucleation typically leads directly to fracture, indicating a more pronounced disorder-trapping effect over lattice trapping, which operates on a larger length and time scale.
%\pj{
%In experiments, common methods for measuring the strength\,\cite{akinwande2017review} of 2D materials include nanoindentation\,\cite{lee2008measurement,bertolazzi2011stretching} and tensile testing\,\cite{feng2022experimentally}. In nanoindentation, a sharp indenter (usually diamond) is pressed into the material's surface to assess its mechanical response under loading. However, this technique poses the potential for surface damage and complicates the interpretation of penetration-induced fracture behavior. In tensile testing, 2D material samples are suspended and stretched using micro-electromechanical systems (MEMS)-based devices to record stress-strain responses until failure. Nonetheless, this method presents challenges in sample preparation and requires precise alignment to prevent buckling.
%}

The edges cleaved by fracture display distinct characteristics across the three representative glassy structures.
In NCG, cracks typically propagate through large crystalline grains ahead of the crack tip and cleave crystalline facets, as the energy cost of crack deflection is relatively high.
In contrast, the smaller and more distorted crystalline grains in Para hinder straight crack propagation, resulting in a more tortuous path.
Conversely, the highly disordered CRN structures facilitate a relatively smooth fracture path.
%with a \AA ngstrom-level roughness.
However, the contrast in the strengths (Fig.\,\ref{Fig_3}a) and edge roughness (Fig.\,\ref{Fig_3}b) of NCG, Para, and CRN structures are not significant in our simulations, suggesting a weak nanostructuring effect.
\rev{The reportedly strong toughening effects observed in experiments may thus be attributed to the chemical heterogeneity in the crystalline and glassy phases and an enhanced 3D nanostructuring effect compared to 2D\cite{gallo2022fracture,tang2023toughening}, which awaits further exploration.}
%\pj{Moreover, the dimension may also play a important role in fracture process of materials, especially in glasses with different microstructures, such as nanopores, crystalline phases and microcracks.
%In 3D glass, the crack tip forms a line, contrasting with the point-like crack tip in 2D glass. 
%This distinction means that in bulk glass, which consists of various microstructures, the linear nature of the crack tip allows interactions with multiple microstructures, leading to possible divergence in several directions. 
%Such divergence could result in crack branching or the formation of curved cracks, thereby increasing energy dissipation.
%To the best of my knowledge, due to limited research on fracture of 2D glass, comparisons between fractures in two- and three-dimensional glass systems are underdocumented, marking a valuable direction for future research. However, this requires more comprehensive and precise force fields that can simultaneously simulate fractures in both two-dimensional and three-dimensional glass systems, where MLFF can become a useful tools.
%}
%By forming atomistically smooth facets, the local roughness of edges cleaved from the crystalline domain is low, \zp{but their kinking features often result in significant roughness in the ceystals}.

\begin{figure*}[h]% [\sidecaptionrelwidth]%[ht]%
{
\includegraphics[width=\textwidth]{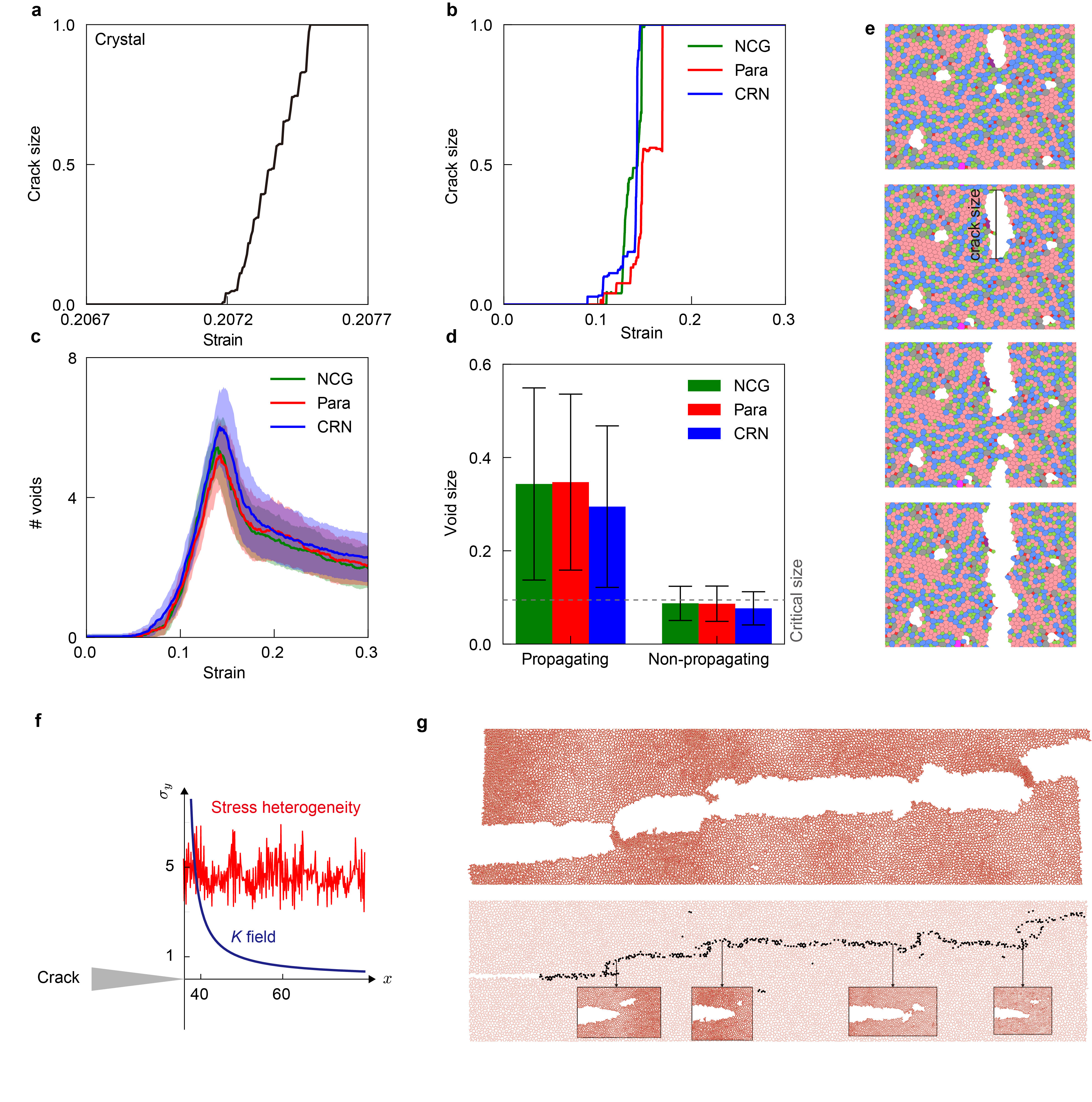}
\caption{}
\label{Fig_4}
}
\end{figure*}

\rev{\subsection{Fracture kinetics and dynamical effects}}
%The propagation of fracture across the simulated samples is studied by extending the simulation time.
There is a signature difference in the fracture kinetics and patterns in crystals and glasses.
In crystals, an advancing crack leaves atomistically smooth crystalline facets behind, although lattice kinks joining these facets in chiral edges can lead to roughening at a length scale larger than the lattice constants.
In contrast, in all three amorphous structures, the cracks become ensnared within the disordered phase, where the cleaved edges show roughness at the length scale of a single Si-O-Si bond (Fig.\,\ref{Fig_4}b). 
%The time and length scales of disorder trapping are much larger than the lattice trapping effect in crystals.
%This kinetic effect may also be crucial for the dynamical fracture processes and can be further explored by studying the loading rate dependence in simulations and experiments.

Crack propagation is facilitated by void nucleation and coalescence in the amorphous phases.
In our AQS simulations, the number of voids increases first by their nucleation and then declines by aggregation (Fig.\,\ref{Fig_4}c).
Propagating cracks are predominantly characterized by the relatively large voids within the samples (Fig.\,\ref{Fig_4}d, e).
\rev{Following the Griffith theory, we find that the critical crack size is $c_{\rm cr} = \frac{4\gamma}{\pi E \varepsilon^2} = 15.08$ \AA, where the values of $\gamma$ and $E$ are the edge energy densities and Young's modulus.
The value of $E$ is extracted from the stress-strain relation of the CRN glass (Fig.\,\ref{Fig_3}a), and $\gamma$ is the edge energy density calculated for 2D crystalline silica that is the lower bound of the value for glasses.}
%which is similar for crystals and glasses
This critical value well separates the propagating and non-propagating voids, validating the linear elastic fracture mechanics theory at the \AA ngstrom scale.
It should be noted that using $\gamma$ instead of fracture toughness in the estimation excludes the non-equilibrium process during fracture\cite{shi2024non}, thus undervaluing $c_{\rm cr}$.
The stress field ahead of the crack tip, averaged over several samples, follows the $r^{-1/2}$ singularity.
However, significant structural heterogeneity in each sample dominates the $K$ field, and may ultimately determine the fracture strength\cite{murali2011atomic} (Fig.\,\ref{Fig_4}f).
Voids can thus nucleate at the weak spots that are not necessarily at the crack tip, and the advancing of cracks can be guided by voids nucleated in front of the crack tip and along its path.
%Fracture and failure of the sample then proceed with void coalescence and bridging of short cracks along the paths.

\rev{Dynamic fracture simulations are conducted to explore the additional inertial effects.
A large sample of $100\times50$ nm$^{2}$ is stretched at strain rates ranging from $10^{-5}$ to $10^{-3}$ ps$^{-1}$ at $0.1$ K.
The Nos\'e-Hoover thermostat with a damping constant of $1$ ps is used.
At a low strain rate of $10^{-5}$ ps$^{-1}$, fracture proceeds by edge cleavage and lattice trapping and edge cleavage, which are guided by voids nucleated ahead of the crack tip, defining the crack paths (Fig.\,\ref{Fig_4}g, Movie S1).
Fragmentation intensifies by void nucleation across the sample at higher strain rates (Fig. S8, Movie S2).
These phenomena have been supported by numerous simulations\cite{ding2015brittle,zhang2022fracture,wang2016nanoductility} and experimental studies\cite{celarie2003glass,shen2021observation,prades2005nano} of glassy structures.
The kinetic and dynamical effects elucidated here through atomistic simulations provide further insights into the mechanisms of failure.}

\section{Conclusion}
\rev{We discussed the strength and fracture behaviors of 2D silica by developing and using a neural network force field for fracture (NN-F$^{3}$) to simulate the failure process, which features chemical accuracy and the capability to model problems of relatively large size, such as fracture.
We find that the strength of 2D glasses is determined by the heterogeneous residual stress, and can be reasonably measured by the local virials or bond lengths in the equilibrium state.
Beyond this strength, voids are nucleated in the amorphous network but do not always trigger fracture propagation.
Only those beyond a critical size defined by the Griffith theory can advance and proceed with continuous coalescence and bridging events.
The kinetics of fracture propagation is renormalized by a disorder trapping effect in addition to lattice trapping and is often guided by voids nucleated ahead of the crack tip.
These findings enrich our understanding of material failure at the atomic level, where the effects of disorder and structural heterogeneity are highlighted with the chemical complexity precluded by choosing 2D silica in the study.
Further experimental studies, potentially conducted \emph{in situ}, are expected to add more insights into the problem\cite{ly_2017,feng2022experimentally}.}

\section*{Supplementary Material}

The Supplementary Material includes the sketch map of the dataset used to train the neural network force field for fracture (NN-F$^{3}$), technical details and validation of NN-F$^{3}$, and additional simulation results supporting the discussion.

\begin{acknowledgments}
\noindent This study was supported by the National Natural Science Foundation of China through grants 12425201 and 52090032.
The computation was performed on the Explorer 1000 cluster system of the Tsinghua National Laboratory for Information Science and Technology.
\end{acknowledgments}

\section*{Data Availability Statement}
\begin{center}
\renewcommand\arraystretch{1.2}
\begin{tabular}{| >{\raggedright\arraybackslash}p{0.3\linewidth} | >{\raggedright\arraybackslash}p{0.65\linewidth} |}
\hline
\textbf{AVAILABILITY OF DATA} & \textbf{STATEMENT OF DATA AVAILABILITY}\\  
\hline
Data openly available in a public repository that issues datasets with DOIs
&
The data that support the findings of this study are openly available in figshare at http://doi.org/10.6084/m9.figshare.25688664.
\\\hline
\end{tabular}
\end{center}

\nocite{*}
\bibliography{main_text}

\clearpage
\newpage

\section*{Figure Captions}

\noindent \textbf{Fig. 1}
\textbf{a-c}, Atomic-level structures of 2D silica nanocrystalline (NCG, \textbf{a}), paracrystalline (Para, \textbf{b}) and continuous random network (CRN, \textbf{c}) glasses.
The inset in panel \textbf{c} shows the CRN structures reported in experimental studies\cite{buchner2014ultrathin}.
\rev{The rings are colored by their number of edges, as illustrated in panel \textbf{a}.}
The inset in panel \textbf{c} is reprinted from B\"uchner et al., Chemistry: A European Journal 20, 9176–9183 (2014) with the permission of John Wiley \& Sons.
\textbf{d}, Ring distributions in NCG, Para, and CRN.
\textbf{e-f}, Radial distribution functions (RDFs) of computer-generated CRN (\textbf{e}) and experimentally-identified\cite{buchner2014ultrathin} (\textbf{f}) structures.
\textbf{g}, The probability distribution function of structural similarity, $s$, which is defined in the text.
The reference structure is shown in the inset.
\textbf{h, i}, Angle distribution functions (ADFs) of computer-generated CRN (\textbf{h}) and experimentally-identified\cite{buchner2014ultrathin} (\textbf{i}) structures.

\noindent \textbf{Fig. 2}
\rev{\textbf{a}, Virial coefficients ($V_\theta$) along the tensile directions (the arrows) in the undeformed, equilibrium structures.}
The positions of fracture initiation are annotated by the circles.
\textbf{b}, Relation between the smallest value of $V_\theta$ (the most stretched state) and the strain to void nucleation for the $144$-atom and $3456$-atom samples.
\textbf{c}, Cumulative probability density function (CDF) of $V_\theta$ at the sites of void nucleation, from the smallest to the largest.
\textbf{d}, Relation between Si-O-Si bond lengths (the distance between two silicon atoms) in equilibrium structures and the strain to void nucleation.
Data in panels (\textbf{b}) and (\textbf{d}) are fitted by exponential functions.
\textbf{e}, The length distribution of bonds at the breaking event.

\noindent \textbf{Fig. 3}
\rev{\textbf{a}, Stress-strain relations of 2D silica NCG, Para, CRN glasses, and crystals (stretched along the armchair direction) under uniaxial tension.
The peak strain values are marked by vertical dashed lines.
\textbf{b}, Roughness of cleaved edges, which is defined as the root mean square (RMS) of edge profiles measured from their average values.
The large standard deviation (SD) in roughness reflects the nanostructuring effect as indicated in the simulation snapshots (panels \textbf{e}, \textbf{h}, and \textbf{k}).
\textbf{c-k}, Simulation snapshots of the undeformed (equilibrium) (\textbf{c, f, i}), void-nucleation (\textbf{d, g, j}), post-fracture (\textbf{e, h, k}) states of NCG (\textbf{c-e}), Para (\textbf{f-h}), and CRN (\textbf{i-k}) structures.}

\noindent \textbf{Fig. 4}
\textbf{a}, Crack size measured in the unit of sample size, which evolves with the simulation time and strain.
\textbf{b}, Relation between strain and the reduced crack size.
\textbf{c}, Relation between strain and the number of voids, which is collected from $100$ simulations of each model.
The shadowed regions are bounded by the SDs.
\textbf{d}, The size of propagating and non-propagating voids, measured in the unit of sample size.
The critical crack size, $c_{\rm cr} = 0.092$ (the dashed line), is calculated from the Griffith theory ($c_{\rm cr} = \frac{4\gamma}{\pi E \varepsilon^2}$), where the values of $\gamma$ and $E$ are the edge energy densities and Young's modulus extracted from our NN-F$^{3}$ simulations, respectively.
$\epsilon$ is the applied strain.
\textbf{e}, Snapshots of crack propagation across a simulated sample showing coalescence and bridging events.
\textbf{f}, $r^{-1/2}$-singular $K$ field at the crack tip and stress heterogeneity measured stress $\sigma_{yy}$ for the CRN glass.
\textbf{g}, Dynamical fracture patterns in the CRN structures showing the disorder-trapping effect and void-guided crack paths.
The black dots represent the cleaved edges and nucleated voids during the fracture process, where the void-guided processes are highlighted in the insets.

\clearpage
\newpage

\clearpage
\newpage

\clearpage
\newpage

\clearpage
\newpage

\end{document}